\documentclass[prd,showpacs,showkeys,nofootinbib,twocolumn]{revtex4-1}

\usepackage{graphicx}
\usepackage{amsmath,amsfonts,amssymb}

\begin{document}

\def\ut{{\underset {\widetilde{\ \ }}u}}
\def\at{{\underset {\widetilde{\ \ }}a}}
\def\ap{\check{a}}

\title{Gravitational clock compass in General Relativity}

\author{Dirk Puetzfeld}
\email{dirk.puetzfeld@zarm.uni-bremen.de}
\homepage{http://puetzfeld.org}
\affiliation{University of Bremen, Center of Applied Space Technology and Microgravity (ZARM), 28359 Bremen, Germany} 

\author{Yuri N. Obukhov}
\email{obukhov@ibrae.ac.ru}
\affiliation{Theoretical Physics Laboratory, Nuclear Safety Institute, 
Russian Academy of Sciences, B.Tulskaya 52, 115191 Moscow, Russia} 

\author{Claus L\"ammerzahl}
\email{claus.laemmerzahl@zarm.uni-bremen.de}
\affiliation{University of Bremen, Center of Applied Space Technology and Microgravity (ZARM), 28359 Bremen, Germany{}}
\affiliation{Institute of Physics, University of Oldenburg, 26111 Oldenburg, Germany}  

\date{ \today}

\begin{abstract}
We show how a suitably prepared set of clocks can be used to determine all components of the gravitational field in General Relativity. We call such an experimental setup a clock compass, in analogy to the usual gravitational compass. Particular attention is paid to the construction of the underlying reference frame. Conceptual differences between the clock compass and the standard gravitational compass, which is based on the measurement of the mutual accelerations between the constituents of a swarm of test bodies, are highlighted. Our results are of direct operational relevance for the setup of networks of clocks, for example in the context of relativistic geodesy.
\end{abstract}


\maketitle


\section{Introduction}\label{introduction_sec}

The question of how the gravitational field can be determined in an operational way, is of fundamental importance in gravitational physics. In this paper, we demonstrate how clocks may be used in a general relativistic context.

In \cite{Puetzfeld:Obukhov:2016:1} we derived a generalized deviation equation by employing the covariant expansion technique based on Synge's world function \cite{Synge:1960,DeWitt:Brehme:1960}. In particular we showed, how the deviation equation, and one of its generalizations, can be used to measure the curvature -- i.e.\ the gravitational field -- by monitoring the mutual accelerations between the constituents of a swarm of test bodies. This led to explicit prescriptions for the setup of the constituents of a device called a ``gravitational compass'' \cite{Szekeres:1965}, i.e.\ a realization of a gradiometer in the context of the theory of General Relativity. 

On the experimental side, modern clocks reached an unprecedented level of accuracy and stability \cite{Chou:etal:2010,Huntemann:etal:2012,Guena:etal:2012,Falke:etal:2014,Bloom:etal:2013,Schioppo:etal:2016} in recent years. An application of clocks for the determination of the gravitational field represents an interesting issue. In analogy with our previous investigation \cite{Puetzfeld:Obukhov:2016:1}, such an ensemble or network of suitably prepared clocks may also be called a clock compass, or a clock gradiometer. In this work, we show how an ensemble of clocks can be used to determine the gravitational field from the mutual frequency comparison of the clocks. 

The structure of the paper is as follows: In section \ref{reference_frames_sec} we work out a suitable set of coordinates which allows for the description of events in the vicinity of a world line. In section \ref{behavior_clocks_sec} we show how an ensemble of clocks has to be prepared to find physical quantities, such as acceleration and velocity, through mutual frequency comparisons of the clocks. In particular, in Sec.~\ref{sec_clock_compass} we determine explicit configurations for clock gradiometers which allow for a measurement of all independent components of the curvature tensor. We draw our conclusions in section \ref{sec_conclusions}. The appendix contains a brief overview of the notations and conventions used throughout the article.

\section{Reference frame: inertial and gravitational effects}\label{reference_frames_sec}

Our previous work \cite{Puetzfeld:Obukhov:2016:1} on the gravitational compass based on deviation equations made clear that a suitable choice of coordinates is crucial for the successful determination of the gravitational field. In particular, the operational realization of the coordinates is of importance when it comes to actual measurements. 

From an experimentalists perspective so-called (generalized) Fermi coordinates appear to be realizable operationally. There have been several suggestions for such coordinates in the literature in different contexts \cite{Fermi:1922:1:2:3,Fermi:1962:US,Veblen:1922,Veblen:Thomas:1923,Synge:1931,Walker:1932,Synge:1960,Manasse:Misner:1963,MTW:1973,Ni:1977,Mashhoon:1977,Ni:Zimmermann:1978,Li:Ni:1978,Ni:1978,Li:Ni:1979,Li:Ni:1979:1,Ashby:etal:1986,Eisele:1987,Fukushima:1988,Semerak:1993,Marzlin:1994:1,Chicone:Mashhoon:2004:1,Bini:etal:2005,Chicone:Mashhoon:2006:1,Chicone:Mashhoon:2006:2,Klein:Collas:2008:1,Klein:Collas:2010:2,Delva:etal:2012,Turyshev:etal:2012}, for a time line of the corresponding research see table~\ref{tab_timeline_coordinates}. In the following we are going to derive the line element in the vicinity of a world line, representing an observer in an arbitrary state of motion, in generalized Fermi coordinates.

\begin{table}
\caption{\label{tab_timeline_coordinates}Time line of works on coordinates.}
\begin{ruledtabular}
\begin{tabular}{llccc}
&&\multicolumn{2}{c}{{Spacetime}}&\\
Year & & Curved & Flat& Acceleration \\
\hline
&&&&\\
1922& Fermi \cite{Fermi:1922:1:2:3,Fermi:1962:US}&&x&x\\
1932& Walker \cite{Walker:1932}&x&&x\\
1960& Synge \cite{Synge:1960}&x&&x\\
1963& Manasse \& Misner \cite{Manasse:Misner:1963}&x&&\\
1973& Misner et al.\ \cite{MTW:1973}&x&&\\
1977& Ni \cite{Ni:1977}&&x&x\\
1977& Mashhoon \cite{Mashhoon:1977}&x&&x\\
1978& Ni \& Zimmermann \cite{Ni:Zimmermann:1978}&x&&x\\
1978& Li \& Ni \cite{Li:Ni:1978} &&x&x\\
1978& Ni \cite{Ni:1978} &x&&\\
1979& Li \& Ni \cite{Li:Ni:1979}&x&&x\\
2004& Chicone \& Mashhoon \cite{Chicone:Mashhoon:2004:1} &&x&x\\
2008& Klein \& Collas \cite{Klein:Collas:2008:1}&x&&x\\
2012& Delva \& Angonin \cite{Delva:etal:2012}&x&&x\\
&&&&\\
\hline
\multicolumn{5}{c}{{Results in special backgrounds (PN, Kerr, etc.)}}\\
\hline
&&&&\\
1986& Ashby \& Bertotti \cite{Ashby:etal:1986}&x&&x\\
1988& Fukushima \cite{Fukushima:1988}&x&&x\\
1993& Semer\'ak \cite{Semerak:1993} &x&&x\\
1994& Marzlin \cite{Marzlin:1994:1} &x&&x\\
2005& Bini et al.\ \cite{Bini:etal:2005}&x&&x\\
2005& Chicone \& Mashhoon \cite{Chicone:Mashhoon:2005:1} &x&x&x\\
2006& Chicone \& Mashhoon \cite{Chicone:Mashhoon:2006:1,Chicone:Mashhoon:2006:2} &x&&\\
2010& Klein \& Collas \cite{Klein:Collas:2010:2}&x&&x\\
2012& Turyshev et al. \cite{Turyshev:etal:2012}&&x&x\\
&&&&\\
\end{tabular}
\end{ruledtabular}
\end{table}

\subsection{Fermi normal coordinates}\label{fermi_subpara}

Following \cite{Veblen:Thomas:1923} we start by taking successive derivatives of the usual geodesic equation. This generates a set of equations of the form (for $n \geq 2$) 
\begin{eqnarray}
\frac{d^n x^a}{ds^n} &=& - \Gamma_{b_1 \dots b_{n}}{}^a \, \frac{d x^{b_1}}{ds} \cdots \frac{d x^{b_{n}}}{ds},
\label{geoddiffset}
\end{eqnarray}
where the $\Gamma$ objects with $n \geq 3$ lower indices are defined by the recurrent relation
\begin{eqnarray}
\Gamma_{b_{1}\dots b_{n}}{}^a := \partial_{(b_1} \Gamma_{b_2 \dots b_{n})}{}^a - (n-1)\, \Gamma_{c (b_1 \dots b_{n-2}}{}^{a} \, \Gamma_{b_{n-1} b_n)}{}^{c} \nonumber \\ \label{gammadefinition}
\end{eqnarray}
from the components of the symmetric linear connection $\Gamma_{bc}{}^a = \Gamma_{cb}{}^a$. 
A solution $x^a=x^a(s)$ of the geodesic equation may then be expressed as a series
\begin{eqnarray}
x^a &=& \left. x^a  \right|_{0} + s \left. \frac{dx^a}{ds} \right|_{0} + \frac{s^2}{2} \left. \frac{d^2x^a}{ds^2} \right|_{0} +  \frac{s^3}{6} \left. \frac{d^3x^a}{ds^3} \right|_{0} + \cdots \nonumber \\
&=& q^a + s v^a - \frac{s^2}{2} \stackrel{0}{\Gamma}_{bc}{\!}^a \, v^b v^c - \frac{s^3}{6} \stackrel{0}{\Gamma}_{bcd}{\!}^a \, v^b v^c v^d - \cdots\,,\nonumber \\
\end{eqnarray}
where in the last line we used $q^a :=\left. x^a  \right|_{0}$, $v^a := \left. \frac{dx^a}{ds} \right|_{0}$, and $\stackrel{0}{\Gamma}_{\dots}{}^a:=\left. \Gamma_{\dots}{}^a \right|_{0}$ for constant quantities at the point around which the series development is performed.

\begin{figure}
\begin{center}
\includegraphics[width=7.8cm,angle=-90]{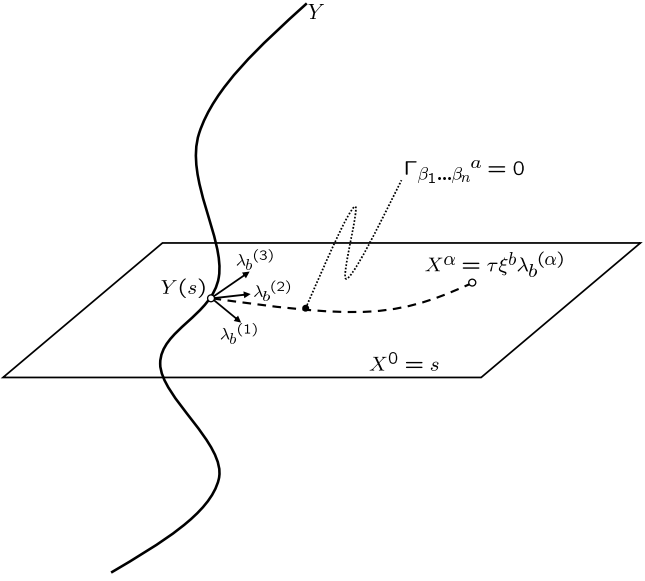}
\end{center}
\caption{\label{fig_1} Construction of the coordinate system around the reference curve $Y$. Coordinates of a point $X$ in the vicinity of $Y(s)$ -- with $s$ representing the proper time along $Y$ -- are constructed by means of a tetrad $\lambda_b{}^{(\alpha)}$. Here $\tau$ is the proper time along the (spacelike) geodesic connecting $Y$ and $X$. By choosing a linear ansatz for the coordinates the derivatives of the connection vanish along the geodesic connecting $Y$ and $X$.}
\end{figure}

Now let us setup coordinates centered on the reference curve $Y$ to describe an adjacent point $X$. For this we consider a unique geodesic connecting $Y$ and $X$. We define our coordinates in the vicinity of a point on $Y(s)$, with proper time $s$, by using a tetrad $\lambda_b{}^{(\alpha)}$ which is Fermi transported along $Y$, i.e.\ 
\begin{eqnarray}
X^0 = s, \quad \quad X^\alpha = \tau \xi^b \lambda_b{}^{(\alpha)}. \label{fermi_ansatz}
\end{eqnarray}
Here $\alpha=1,\dots,3$, and $\tau$ is the proper time along the (spacelike) geodesic connecting $Y(s)$ and $X$. The $\xi^b$ are constants, and it is important to notice that the tetrads are functions of the proper time $s$ along the reference curve $Y$, but independent of $\tau$. See figure \ref{fig_1} for further explanations. By means of this linear ansatz (\ref{fermi_ansatz}) for the coordinates in the vicinity of $Y$, we obtain for the derivatives w.r.t.\ $\tau$ along the connecting geodesic ($n \geq 1$):
\begin{eqnarray}
\frac{d^n X^0}{d \tau^n} &=& 0, \nonumber \\
\frac{d X^\alpha}{d \tau} &=& \xi^b \lambda_b{}^{(\alpha)}, \quad \frac{d^{n+1} X^\alpha}{d \tau^{n+1}} =0 .
\end{eqnarray}
In other words, in the chosen coordinates (\ref{fermi_ansatz}), along the geodesic connecting $Y$ and $X$, one obtains for the derivatives ($n \geq 2$)
\begin{eqnarray}
\Gamma_{b_1 \dots b_{n}}{}^a \, \frac{d X^{b_1}}{d\tau} \cdots \frac{d X^{b_{n}}}{d\tau}=0.
\label{geoddiffset_2}
\end{eqnarray}
This immediately yields 
\begin{eqnarray}
\Gamma_{\beta_1 \dots \beta_{n}}{}^a = 0, \label{connection_deriv_condition}
\end{eqnarray}
along the connecting curve, in the region covered by the linear coordinates as defined above.

The Fermi normal coordinate system cannot cover the whole spacetime manifold. By construction, it is a good way to describe the physical phenomena in a small region around the world line of an observer. The smallness of the corresponding domain depends on the motion of the latter, in particular, on the magnitudes of the acceleration $|a|$ and angular velocity $|\omega|$ of the observer which set the two characteristic lengths: $\ell_{\textrm{tr}} = c^2/|a|$ and $\ell_{\textrm{rot}} = c/|\omega|$. The Fermi coordinate system $X^\alpha$ provides a good description for the region $|X|/\ell \ll 1$. For example, this condition is with a high accuracy valid in terrestrial laboratories since $\ell_{\textrm{tr}} = c^2/|g_\oplus| \approx 10^{16}$m (one light year), and $\ell_{\textrm{rot}} = c/|\Omega_\oplus| \approx 4\times 10^{12}$m (27 astronomical units). Note, however, that for a particle accelerated in a storage ring $\ell \approx 10^{-6}$m. Furthermore, the region of validity of the Fermi coordinate system is restricted by the strength of the gravitational field in the region close to the reference curve, $\ell_{\textrm{grav}} = \textrm{min}\{|R_{abcd}|^{-1/2}, |R_{abcd}|/|R_{abcd,e}|\}$, so that the curvature should have not yet caused geodesics to cross. We always assume that there is a unique geodesic connecting $Y$ and $X$.

\subsection{Explicit form of the connection}

At the lowest order, in flat spacetime, the connection of a noninertial system that is accelerating with $a^\alpha$ and rotating with angular velocity $\omega^\alpha$ at the origin of the coordinate system is
\begin{eqnarray}
\Gamma_{00}{}^{0}&=&\Gamma_{\alpha \beta}{}^{c}=0,\quad
\Gamma_{00}{}^{\alpha} = a^\alpha,\quad \nonumber \\
\Gamma_{0\alpha}{}^{0} &=& a_\alpha, \quad
\Gamma_{0\beta}{}^{\alpha} = - \varepsilon^{\alpha}{}_{\beta \gamma} \omega^\gamma. \label{connection_expl}
\end{eqnarray}
Hereafter $\varepsilon_{\alpha\beta\gamma}$ is the 3-dimensional totally antisymmetric Levi-Civita symbol, and the Euclidean 3-dimensional metric $\delta_{\alpha\beta}$ is used to raise and lower the spatial (Greek) indices, in particular $a_\alpha = \delta_{\alpha\beta}a^\beta$ and $\varepsilon^{\alpha}{}_{\beta \gamma} = \delta^{\alpha\delta}\varepsilon_{\delta\beta \gamma}$.
For the time derivatives we have
\begin{eqnarray}
\partial_{0}\Gamma_{00}{}^{0}&=&\partial_{0} \Gamma_{\alpha \beta}{}^{c}=0,\quad
\partial_{0} \Gamma_{00}{}^{\alpha} = \partial_{0} a^\alpha =: b^{\alpha},\quad \nonumber \\
\partial_{0} \Gamma_{0\alpha}{}^{0} &=& b_{\alpha}, \,
\partial_{0} \Gamma_{0\beta}{}^{\alpha} = - \varepsilon^{\alpha}{}_{\beta \gamma} \partial_{0} \omega^\gamma =: - \varepsilon^{\alpha}{}_{\beta \gamma} \eta^\gamma. \label{0_connection_expl}
\end{eqnarray}
From the definition of the curvature we can express the next order of derivatives of the connection in terms of the curvature:
\begin{eqnarray}
\partial_{\alpha}\Gamma_{00}{}^{0} &=& b_\alpha - a_\beta \varepsilon^{\beta}{}_{\alpha \gamma} \omega^\gamma, \nonumber\\
\partial_{\alpha}\Gamma_{00}{}^{\beta}&=& -\,R_{0 \alpha 0}{}^\beta-\varepsilon^\beta{}_{\alpha \gamma} \eta^{\gamma} + a_\alpha a^\beta - \delta_\alpha^\beta \omega_\gamma \omega^\gamma + \omega_\alpha \omega^\beta, \nonumber \\
\partial_{\alpha}\Gamma_{0\beta}{}^{0}&=& -\,R_{0 \alpha \beta}{}^0 - a_\alpha a_\beta , \nonumber \\
\partial_{\alpha}\Gamma_{0\beta}{}^{\gamma}&=& -\,R_{0 \alpha \beta}{}^\gamma + \varepsilon^\gamma{}_{\alpha \delta} \omega^\delta a_\beta. \label{1_connection_expl}
\end{eqnarray}

Using (\ref{connection_deriv_condition}), we derive the spatial derivatives 
\begin{eqnarray}
\partial_{\alpha}\Gamma_{\beta \gamma}{}^{d}&=& \frac{2}{3} R_{\alpha (\beta \gamma)}{}^d, \label{1_connection_expl_1} 
\end{eqnarray}
see also the general solution given in the appendix B of \cite{Puetzfeld:Obukhov:2016:1}.

\subsection{Explicit form of the metric}

In order to determine, in the vicinity of the reference curve $Y$, the form of the metric at the point $X$ in coordinates $y^a$ centered on $Y$, we start again with an expansion of the metric around the reference curve
\begin{eqnarray}
\left. g_{ab} \right|_{X} &=& \left. g_{ab}  \right|_{Y} + \left. g_{ab,c} \right|_{Y} y^c + \frac{1}{2} \left. g_{ab,cd} \right|_{Y} y^c y^d  + \cdots\,. \nonumber \\
\end{eqnarray}
Of course in normal coordinates we have $\left. g_{ab}  \right|_{Y} = {\eta}_{ab}$, whereas the derivatives of the metric have to be calculated, and the result actually depends on which type of coordinates we want to use. The derivatives of the metric may be expressed just by successive differentiation of the metricity condition $\nabla_c g_{ab} =0$:
\begin{eqnarray}
g_{ab,c} &=& 2 \, g_{d(a} \Gamma_{b)c}{}^d, \nonumber \\
g_{ab,cd} &=& 2 \, \left( \partial_d g_{e(a} \Gamma_{b)c}{}^e + \partial_d \Gamma_{c(a}{}^e  g_{b)e} \right), \nonumber \\
&\vdots&  \label{metricity_cond_explicit}
\end{eqnarray}
In other words, we can iteratively determine the metric by plugging in the explicit form of the connection and its derivatives from above.

In combination with (\ref{connection_expl}) one finds:
\begin{eqnarray}
g_{00,0} &=& g_{0\alpha,0} = g_{\alpha \beta,0} = g_{\alpha \beta, \gamma}  = 0, \nonumber \\
g_{00,\alpha} &=& 2 a_\alpha, \quad g_{0 \alpha , \beta} = \varepsilon_{\alpha \beta \gamma} \omega^\gamma. \label{1_metric_explicit}
\end{eqnarray}

For the second-order derivatives of the metric we obtain, again using (\ref{metricity_cond_explicit}) in combination with (\ref{0_connection_expl}), (\ref{1_connection_expl}), and (\ref{1_metric_explicit}):
\begin{eqnarray}
g_{00,00} &=& g_{0\alpha,0 0} = g_{\alpha \beta,00}= g_{\alpha \beta,\gamma 0} = 0, \quad g_{00,\alpha 0} = 2 b_\alpha, \nonumber \\
g_{0\alpha,\beta 0} &=& - \varepsilon^{\gamma}{}_{\beta \delta} \eta^{\delta} g_{\alpha \gamma} = \varepsilon_{\alpha \beta \gamma} \eta^{\gamma}, \nonumber \\
g_{00,\alpha \beta} &=& -\,2 R_{0\beta \alpha}{}^0 + 2 a_\alpha a_\beta - 2 \delta_{\alpha \beta} \omega_\gamma \omega^\gamma + 2 \omega_\alpha \omega_\beta ,\nonumber \\
g_{0\alpha,\beta \gamma} &=&  -\,{\frac 43} R_{\alpha(\beta\gamma)}{}^0,\quad
g_{\alpha \beta, \gamma \delta} = {\frac 23} R_{\gamma (\alpha \beta) \delta}. \label{2_metric_explicit}
\end{eqnarray}
Note that $R_{0\beta \alpha}{}^0 + R_{\alpha 0\beta}{}^0 + R_{\beta \alpha 0}{}^0 \equiv 0$, in view of the Ricci identity. Since $R_{\beta \alpha 0}{}^0 = 0$, we thus find $R_{0\beta \alpha}{}^0 = R_{0(\beta \alpha)}{}^0$. 

As a result, we derive the line element in the Fermi coordinates (up to the second order):
\begin{widetext}
\begin{eqnarray}
\left. ds^2 \right|_X(y^{0},y^{\alpha}) &=& (dy^{0})^2 \bigr[1 + 2 a_{\alpha} y^{\alpha} + 2 b_\alpha y^\alpha y^0 + (a_\alpha a_\beta - \delta_{\alpha \beta} \omega_\gamma \omega^\gamma + \omega_\alpha \omega_\beta - R_{0\alpha\beta 0}) y^\alpha y^\beta \bigr]    \nonumber \\ 
&&+ 2 dy^{0} dy^{\alpha} \bigr[ \varepsilon_{\alpha \beta \gamma} \omega^\gamma y^\beta + \varepsilon_{\alpha \beta \gamma} \eta^\gamma y^\beta y^0 - \frac{2}{3} R_{\alpha \beta \gamma 0} y^{\beta} y^{\gamma} \bigr] \nonumber \\
&&-dy^{\alpha} dy^{\beta}\bigr[\delta_{\alpha\beta} - \frac{1}{3} R_{\gamma \alpha \beta \delta} y^{\gamma} y^{\delta} \bigr] + {\mathcal O}(3).
\label{fermi_normal_ds}
\end{eqnarray}
It is worthwhile to notice that we can recast this result as
\begin{eqnarray}
\left. ds^2 \right|_X(y^{0},y^{\alpha}) &=& (dy^{0})^2 \bigr[1 + 2 \overline{a}_{\alpha} y^{\alpha} + (\overline{a}_\alpha \overline{a}_\beta - \delta_{\alpha \beta} \overline{\omega}_\gamma \overline{\omega}^\gamma + \overline{\omega}_\alpha \overline{\omega}_\beta - R_{0\alpha\beta 0}) y^\alpha y^\beta \bigr]    \nonumber \\ 
&&+ 2 dy^{0} dy^{\alpha} \bigr[ \varepsilon_{\alpha \beta \gamma} \overline{\omega}^\gamma y^\beta - \frac{2}{3} R_{\alpha \beta \gamma 0} y^{\beta} y^{\gamma} \bigr] -dy^{\alpha} dy^{\beta}\bigr[\delta_{\alpha\beta} - \frac{1}{3} R_{\gamma \alpha \beta \delta} y^{\gamma} y^{\delta} \bigr] + {\mathcal O}(3),
\label{fermi_normal_ds1}
\end{eqnarray}
by introducing $\overline{a}_{\alpha} = a_\alpha + y^0\partial_0a_\alpha = a_\alpha + y^0 b_\alpha$ and $\overline{\omega}_{\alpha} = \omega_\alpha + y^0\partial_0\omega_\alpha = \omega_\alpha + y^0 \eta_\alpha$ which represent the power expansion of the time dependent acceleration and angular velocity. 

\end{widetext}

\section{Apparent behavior of clocks}\label{behavior_clocks_sec}

The results from the last section may now be used to describe the behavior of clocks in the vicinity of the reference world line, around which the coordinates were constructed. 

There is one interesting peculiarity about writing the metric like in (\ref{fermi_normal_ds}), i.e.\ one obtains clock effects which depend on the acceleration of the clock (just integrate along a curve in those coordinates and the terms with $a$ and $\omega$ will of course contribute to the proper time along the curve). This behavior of clocks is of course due to the choice of the noninertial observer, and they are {\it only} present along curves which do not coincide with the observers world line. Recall that, by construction, one has Minkowski's metric along the world line of the observer, which is also the center of the coordinate system in which (\ref{fermi_normal_ds}) is written -- all inertial effects vanish at the origin of the coordinate system. 

\subsection{Flat case}\label{paragraph_fermi_normal_flat_freq_ratio}

We start with the flat spacetime and switch to a quantity which is directly measurable, i.e.\ the proper time quotient of two clocks located at $Y$ and $X$. It is worthwhile to note that for a flat spacetime, $R_{ijk}{}^l = 0$, the interval (\ref{fermi_normal_ds}) reduces to the Hehl-Ni \cite{Hehl:Ni:1990} line element of a noninertial (rotating and accelerating) system:
\begin{widetext}
\begin{eqnarray}
\left.ds^2 \right|_X(y^{0},y^{\alpha}) = (1 + \overline{a}_\alpha y^\alpha)^2(dy^{0})^2 - \delta_{\alpha\beta} (dy^{\alpha} + \varepsilon^\alpha{}_{\mu\nu}\overline{\omega}^\mu y^\nu\,dy^0)
(dy^{\beta} + \varepsilon^\beta{}_{\rho\sigma}\overline{\omega}^\rho y^\sigma\,dy^0) + {\mathcal O}(3),\nonumber\\
\label{fermi_normal_ds2}
\end{eqnarray}
From (\ref{fermi_normal_ds}) we derive
\begin{eqnarray}
\left(\frac{ds|_X}{ds|_Y}\right)^2 &=&\left(\frac{dy^{0}}{ds|_Y}\right)^2 \bigr[ 1 - \delta_{\alpha \beta} v^\alpha v^\beta + 2 a_\alpha y^\alpha + 2 b_\alpha y^\alpha y^0 + y^\alpha y^\beta \left( a_\alpha a_\beta - \delta_{\alpha \beta} \omega_\gamma \omega^\gamma + \omega_\alpha \omega_\beta \right) \nonumber \\
&&+ 2 v^\alpha \varepsilon_{\alpha \beta \gamma}  \left(y^\beta \omega^\gamma + y^0 y^\beta \eta^\gamma \right) \bigr] + {\mathcal O}(3) \\
&=&1+ \frac{1}{1 - \delta_{\alpha \beta} v^\alpha v^\beta } \bigr[ 2 a_\alpha y^\alpha + 2 b_\alpha y^\alpha y^0  +  y^\alpha y^\beta \left( a_\alpha a_\beta - \delta_{\alpha \beta} \omega_\gamma \omega^\gamma + \omega_\alpha \omega_\beta \right) \nonumber \\
&&+ 2 v^\alpha \varepsilon_{\alpha \beta \gamma}  \left(y^\beta \omega^\gamma + y^0 y^\beta \eta^\gamma \right) \bigr] + {\mathcal O}(3). \label{fermi_normal_flat_freq_ratio}
\end{eqnarray}
Here we introduced the velocity $v^\alpha = dy^\alpha/dy^0$. Defining 
\begin{equation}
V^\alpha := v^\alpha + \varepsilon^\alpha{}_{\beta \gamma}\overline{\omega}^\beta y^\gamma,\label{Va}
\end{equation}
we can rewrite the above relation more elegantly as
\begin{eqnarray}
\left(\frac{ds|_X}{ds|_Y}\right)^2 &=&\left(\frac{dy^{0}}{ds|_Y}\right)^2 \bigr[ (1 + \overline{a}_\alpha y^\alpha)^2 - \delta_{\alpha \beta} V^\alpha V^\beta \bigr] + {\mathcal O}(3). \nonumber\\
&& \label{fermi_normal_flat_freq_ratio1}
\end{eqnarray}

Equation (\ref{fermi_normal_flat_freq_ratio}) is reminiscent of the situation which we encountered in case of the gravitational compass, i.e.\ we may look at this measurable quantity depending on how we prepare the 
\begin{eqnarray}
C\left( y^\alpha, y^0 , v^\alpha , a^\alpha, \omega^\alpha, b^\alpha, \eta^\alpha \right) := \left(\frac{ds|_X}{ds|_Y}\right)^2. \label{general_freq_ratio_definition}
\end{eqnarray}

\subsection{Curved case}\label{paragraph_fermi_normal_curved_freq_ratio}

Now let us investigate the curved spacetime, after all we are interested in mapping the gravitational field by means of clock comparison. The frequency ratio becomes: 
\begin{eqnarray}
\left(\frac{ds|_X}{ds|_Y}\right)^2 &=&1+ \frac{1}{1 - \delta_{\alpha \beta} v^\alpha v^\beta } \bigr[ 2 a_\alpha y^\alpha + 2 b_\alpha y^\alpha y^0  +  y^\alpha y^\beta \left( a_\alpha a_\beta - R_{0 \alpha \beta 0} - \delta_{\alpha \beta} \omega_\gamma \omega^\gamma + \omega_\alpha \omega_\beta \right)  \nonumber \\
&& + 2 v^\alpha  \varepsilon_{\alpha \beta \gamma} \left(  y^\beta \omega^\gamma +  y^0 y^\beta \eta^\gamma \right) - \frac{4}{3} v^\alpha y^\beta y^\gamma R_{ \alpha \beta \gamma 0} + \frac{1}{3} v^\alpha v^\beta y^\gamma y^\delta R_{\gamma \alpha \beta \delta} \bigr] + {\mathcal O}(3) .
\label{fermi_normal_curved_freq_ratio}
\end{eqnarray}
Analogously to the flat case in (\ref{general_freq_ratio_definition}), we introduce a shortcut for the measurable frequency ratio in a curved background, denoting its dependence on different quantities as $C\left( y^\alpha, y^0 , v^\alpha , a^\alpha, \omega^\alpha, b^\alpha, \eta^\alpha, R_{\alpha \beta \gamma \delta} \right)$.

Note that in the flat, as well as in the curved case, the frequency ratio becomes independent of $b^\alpha$ and $\eta^\alpha$ on the three-dimensional slice with fixed $y^0$ (since we can always choose our coordinate time parameter $y^0=0$), i.e.\ we have $C\left( y^\alpha, v^\alpha , a^\alpha, \omega^\alpha \right)$ and $C\left( y^\alpha, v^\alpha , a^\alpha, \omega^\alpha, R_{\alpha \beta \gamma \delta} \right)$ respectively. 
\end{widetext}

\section{Clock compass }\label{sec_clock_compass}

We now consider different setups of clocks to measure physical quantities by means of mutual frequency comparisons. For example, we could ask the question: can we detect rotation just by clock comparison, i.e.\ can we measure all three components of $\omega^{\alpha}$, by a suitable setup of clocks w.r.t.\ to the clock on our reference world line $Y$? 

Here our strategy is similar to our analysis of the gravitational compass in \cite{Puetzfeld:Obukhov:2016:1}. 
We start by labeling different initial values for the clocks: 
\begin{eqnarray}
&&{}^{(1)}y^{{\alpha}}=\left(\begin{array}{c} 1 \\ 0\\ 0\\ \end{array} \right),
{}^{(2)}y^{{\alpha}}=\left(\begin{array}{c} 0 \\ 1\\ 0\\ \end{array} \right),
{}^{(3)}y^{{\alpha}}=\left(\begin{array}{c} 0 \\ 0\\ 1\\ \end{array} \right), \nonumber \\
&&{}^{(4)}y^{{\alpha}}=\left(\begin{array}{c} 1\\ 1\\ 0\\ \end{array} \right),
{}^{(5)}y^{{\alpha}}=\left(\begin{array}{c}  0\\ 1\\ 1\\ \end{array} \right),
{}^{(6)}y^{{\alpha}}=\left(\begin{array}{c}  1\\ 0\\ 1\\ \end{array} \right), \label{position_setup}
\end{eqnarray}
and
\begin{eqnarray}
&&{}^{(1)}v^{{\alpha}}=\left(\begin{array}{c} c_{11} \\ 0\\ 0\\ \end{array} \right),
{}^{(2)}v^{{\alpha}}=\left(\begin{array}{c} 0 \\ c_{22}\\ 0\\ \end{array} \right),
{}^{(3)}v^{{\alpha}}=\left(\begin{array}{c} 0 \\ 0\\ c_{33}\\ \end{array} \right), \nonumber \\
&&{}^{(4)}v^{{\alpha}}=\left(\begin{array}{c} c_{41}\\ c_{42}\\ 0\\ \end{array} \right),
{}^{(5)}v^{{\alpha}}=\left(\begin{array}{c}  0\\ c_{52}\\ c_{53}\\ \end{array} \right),
{}^{(6)}v^{{\alpha}}=\left(\begin{array}{c}  c_{61}\\ 0\\ c_{63}\\ \end{array} \right), \nonumber \\ \label{velocity_setup}
\end{eqnarray}
and
\begin{eqnarray}
&&{}^{(1)}a^{{\alpha}}=\left(\begin{array}{c} d_{11} \\ 0\\ 0\\ \end{array} \right),
{}^{(2)}a^{{\alpha}}=\left(\begin{array}{c} 0 \\ d_{22}\\ 0\\ \end{array} \right),
{}^{(3)}a^{{\alpha}}=\left(\begin{array}{c} 0 \\ 0\\ d_{33}\\ \end{array} \right), \nonumber \\
&&{}^{(4)}a^{{\alpha}}=\left(\begin{array}{c} d_{41}\\ d_{42}\\ 0\\ \end{array} \right),
{}^{(5)}a^{{\alpha}}=\left(\begin{array}{c}  0\\ d_{52}\\ d_{53}\\ \end{array} \right),
{}^{(6)}a^{{\alpha}}=\left(\begin{array}{c}  d_{61}\\ 0\\ d_{63}\\ \end{array} \right), \nonumber \\ \label{linaccel_setup}
\end{eqnarray}
and
\begin{eqnarray}
&&{}^{(1)}\omega^{{\alpha}}=\left(\begin{array}{c} e_{11} \\ 0\\ 0\\ \end{array} \right),
{}^{(2)}\omega^{{\alpha}}=\left(\begin{array}{c} 0 \\ e_{22}\\ 0\\ \end{array} \right),
{}^{(3)}\omega^{{\alpha}}=\left(\begin{array}{c} 0 \\ 0\\ e_{33}\\ \end{array} \right), \nonumber \\
&&{}^{(4)}\omega^{{\alpha}}=\left(\begin{array}{c} e_{41}\\ e_{42}\\ 0\\ \end{array} \right),
{}^{(5)}\omega^{{\alpha}}=\left(\begin{array}{c}  0\\ e_{52}\\ e_{53}\\ \end{array} \right),
{}^{(6)}\omega^{{\alpha}}=\left(\begin{array}{c}  e_{61}\\ 0\\ e_{63}\\ \end{array} \right). \nonumber \\ \label{rotaccel_setup}
\end{eqnarray}
Here the $c$'s, $d$'s and $e$'s are real-valued parameters. 

\subsection{Linear acceleration determination}\label{paragraph_fermi_normal_flat_linear_accel_determination}

\begin{figure}
\begin{center}
\includegraphics[width=3.5cm,angle=-90]{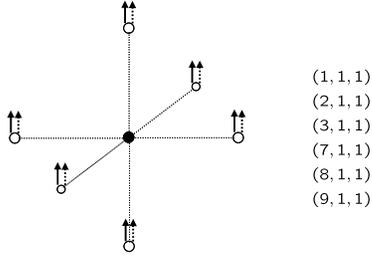}
\end{center}
\caption{\label{fig_2} Symbolical sketch of the explicit solution for linear acceleration (\ref{linaccel_solution_C}). In total 6 suitably prepared clocks (hollow circles) are needed to determine all acceleration components. The observer is denoted by the black circle.}
\end{figure}

Now let us search for a configuration of clocks which allows for a determination of the three components of the linear acceleration $a^\alpha$ of the observer. Assuming that all other quantities can be prescribed by the experimentalist, we rearrange (\ref{general_freq_ratio_definition}) as follows: 
\begin{eqnarray}
2a_\alpha y^\alpha + a_\alpha a_\beta y^\alpha y^\beta = B(y^\alpha, v^\alpha,\omega^\alpha),
\end{eqnarray} 
where all the measured frequency ratios, as well as all prescribed quantities are collected on the rhs
\begin{eqnarray}
B(y^\alpha, v^\alpha,\omega^\alpha) &:=& \left(1 - v^2\right) \left( C-1 \right) - y^\alpha y^\beta \left( \omega_\alpha \omega_\beta - \delta_{\alpha \beta} \omega^2 \right) \nonumber \\
&&- 2 v^\alpha \varepsilon_{\alpha \beta \gamma}  y^\beta \omega^\gamma. \label{rhs_accel_determination}
\end{eqnarray}
Note that for brevity we suppress the functional dependence on parameters of the measured frequency ratios on the rhs. Taking into account (\ref{position_setup})-(\ref{rotaccel_setup}), we end up with the system
\begin{eqnarray}
2a_\alpha {}^{(n)}y^\alpha + a_\alpha a_\beta {}^{(n)}y^\alpha {}^{(n)}y^\beta &=& B({}^{(n)}y^\alpha, {}^{(m)}v^\alpha,{}^{(p)}\omega^\alpha) \nonumber \\
&=& {}^{(n,m,p)}B .
\end{eqnarray}
Inserting eq.\ (\ref{position_setup}) yields the set (redundant equations are not displayed)
\begin{eqnarray}
a_1^2+2a_1&=&{}^{(1,1,1)}B, \label{linaccel_1} \\
a_2^2+2a_2&=&{}^{(2,1,1)}B, \label{linaccel_2} \\ 
a_3^2+2a_3&=&{}^{(3,1,1)}B, \label{linaccel_3} \\
2a_1+2a_2+(a_1+a_2)^2&=&{}^{(4,1,1)}B, \label{linaccel_4} \\
2a_2+2a_3+(a_2+a_3)^2&=&{}^{(5,1,1)}B, \label{linaccel_5} \\
2a_1+2a_3+(a_1+a_3)^2&=&{}^{(6,1,1)}B. \label{linaccel_6}
\end{eqnarray}
This system does not allow for an extraction of the linear accelerations, but this can be achieved by the introduction of clocks at the positions
\begin{eqnarray}
{}^{(7)}y^{{\alpha}}=\left(\begin{array}{c} -1 \\ 0\\ 0\\ \end{array} \right),
{}^{(8)}y^{{\alpha}}=\left(\begin{array}{c} 0 \\ -1\\ 0\\ \end{array} \right),
{}^{(9)}y^{{\alpha}}=\left(\begin{array}{c} 0 \\ 0\\ -1\\ \end{array} \right). \nonumber\\ \label{additional_positions}
\end{eqnarray}
This yields a set of three equations like (\ref{linaccel_1})-(\ref{linaccel_3}), which can be subtracted from each other, leading to
\begin{eqnarray}
a_\alpha=\frac{1}{4}\left({}^{(\alpha,1,1)}B - {}^{(\alpha+6,1,1)}B\right). 
\label{linaccel_solution}
\end{eqnarray}
In terms of the $C$'s, for which we use here and in the following the same shorthand notation as for the $B$'s, we have 
\begin{eqnarray}
a_\alpha=\frac{1}{4} \left(1-c_{11}^2\right) \left({}^{(\alpha,1,1)}C - {}^{(\alpha+6,1,1)}C\right). \label{linaccel_solution_C}
\end{eqnarray}
Hence we need 6 clocks to determine all components of the linear acceleration $a_\alpha$; see figure \ref{fig_2} for a symbolical sketch of the solution.

\subsection{Rotational velocity determination}\label{paragraph_fermi_normal_flat_rotational_velo_determination}

\begin{figure}
\begin{center}
\includegraphics[width=3.2cm,angle=-90]{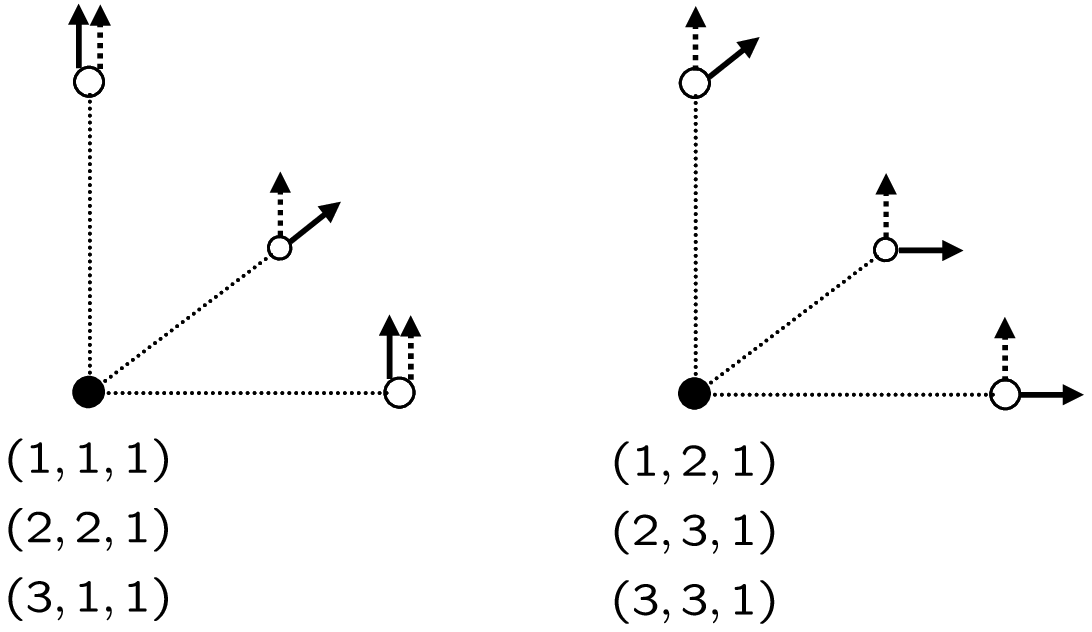}
\end{center}
\caption{\label{fig_3} Symbolical sketch of the explicit solution for the rotational velocity (\ref{rotvel_solution_C}). In total 6 suitably prepared clocks (hollow circles) are needed to determine all velocity components. The observer is denoted by the black circle.}
\end{figure}

Analogously to the strategy in the preceding section, we rearrange the system (\ref{general_freq_ratio_definition}) as follows:
\begin{eqnarray}
 2 v^\alpha \varepsilon_{\alpha \beta \gamma}  y^\beta \omega^\gamma - y^\alpha y^\beta \left(\delta_{\alpha \beta} \omega^2 - \omega_\alpha \omega_\beta \right) = B(y^\alpha, v^\alpha,a^\alpha),\nonumber \\
\end{eqnarray} 
where
\begin{eqnarray}
B(y^\alpha, v^\alpha, a^\alpha) &:=& \left(1 - v^2\right) \left( C-1 \right) - 2a_\alpha y^\alpha - a_\alpha a_\beta y^\alpha y^\beta. \nonumber \\ \label{rhs_accel_determination_2}
\end{eqnarray}
Taking into account (\ref{position_setup})-(\ref{rotaccel_setup}) we end up with
\begin{eqnarray}
&& 2{}^{(m)}v^\alpha \varepsilon_{\alpha \beta \gamma}  {}^{(n)}y^\beta \omega^\gamma - {}^{(n)}y^\alpha {}^{(n)}y^\beta \left(\delta_{\alpha \beta} \omega^2 -\omega_\alpha \omega_\beta \right) \nonumber \\
&=&  B({}^{(n)}y^\alpha, {}^{(m)}v^\alpha,{}^{(p)}a^\alpha) = {}^{(n,m,p)}B .
\end{eqnarray}
Consequently the rotational velocity can be determined with the help of 6 clocks, an explicit solution being
\begin{eqnarray}
\omega^1&=&\frac{1}{2c_{33}} \left({}^{(2,2,1)}B - {}^{(2,3,1)}B\right), \nonumber \\
\omega^2&=&\frac{1}{2c_{11}} \left({}^{(3,3,1)}B - {}^{(3,1,1)}B\right), \nonumber \\
\omega^3&=&\frac{1}{2c_{22}} \left({}^{(1,1,1)}B - {}^{(1,2,1)}B\right), \label{rotvel_solution}
\end{eqnarray}
or explicitly in terms of the $C$'s
\begin{eqnarray}
\omega^1&=&\frac{1-c_{33}^2}{2c_{33}} \left[\frac{1-c_{22}^2}{1-c_{33}^2}\left({}^{(2,2,1)}C - 1 \right)-  {}^{(2,3,1)}C + 1 \right], \nonumber \\
\omega^2&=&\frac{1-c_{11}^2}{2c_{11}} \left[\frac{1-c_{33}^2}{1-c_{11}^2}\left({}^{(3,3,1)}C - 1 \right)-  {}^{(3,1,1)}C + 1 \right], \nonumber \\
\omega^3&=&\frac{1-c_{22}^2}{2c_{22}} \left[\frac{1-c_{11}^2}{1-c_{22}^2}\left({}^{(1,1,1)}C - 1 \right)-  {}^{(1,2,1)}C + 1 \right]. \nonumber \\\label{rotvel_solution_C}
\end{eqnarray}
See figure \ref{fig_3} for a symbolical sketch of the solution.

\subsection{Linear velocity determination}\label{paragraph_fermi_normal_flat_velocity_determination}

Again we rearrange the system (\ref{general_freq_ratio_definition}) as follows:
\begin{eqnarray}
 \left(1 - v^2\right) \left(C-1 \right) - 2 v^\alpha \varepsilon_{\alpha \beta \gamma}  y^\beta \omega^\gamma  = B(y^\alpha, a^\alpha,\omega^\alpha),\nonumber \\
\end{eqnarray} 
where
\begin{eqnarray}
B(y^\alpha, a^\alpha, \omega^\alpha) &:=&  2 a_\alpha y^\alpha + y^\alpha y^\beta \left(a_\alpha a_\beta- \delta_{\alpha \beta} \omega^2 + \omega_\alpha \omega_\beta \right) . \nonumber \\ \label{rhs_velocity_determination}
\end{eqnarray}
This yields
\begin{eqnarray}
&&  \left(1 - \delta_{\alpha \beta}v^\alpha v^\beta\right) \left({}^{(n,m,p)}C-1 \right) - 2 v^\alpha \varepsilon_{\alpha \beta \gamma}  {}^{(n)}y^\beta {}^{(p)}\omega^\gamma \nonumber \\
&& = B({}^{(n)}y^\alpha, {}^{(m)}a^\alpha,{}^{(p)}\omega^\alpha) = {}^{(n,m,p)}B .
\end{eqnarray}
From this system we can determine the linear velocity as follows:
\begin{eqnarray}
v^1&=&\frac{1}{2e_{33}} \left[\left({}^{(2,2,3)}C - 1 \right) A - d_{22}^2+e_{33}^2-2d_{22}\right], \nonumber \\
v^2&=&\frac{1}{2e_{33}} \left[-\left({}^{(1,1,3)}C - 1 \right) A + d_{11}^2-e_{33}^2+2d_{11}\right], \nonumber \\
v^3&=&\frac{1}{2e_{22}} \left[\left({}^{(1,1,2)}C - 1 \right) A - d_{11}^2+e_{22}^2-2d_{11}\right], \label{velocity_solution}
\end{eqnarray}
where the common factor is given by
\begin{eqnarray}
A:=\frac{d_{11}^2+2d_{11}}{{}^{(1,1,1)}C-1}.
\end{eqnarray}

\begin{figure}
\begin{center}
\includegraphics[width=2.1cm,angle=-90]{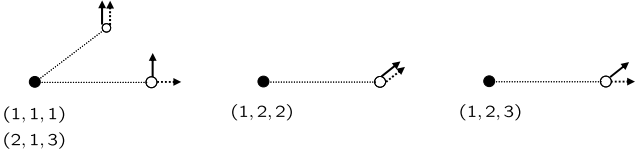}
\end{center}
\caption{\label{fig_4} Symbolical sketch of the explicit solution for the linear velocity (\ref{alternative_velocity_solution}). In total 4 suitably prepared clocks (hollow circles) are needed to determine all velocity components. The observer is denoted by the black circle.}
\end{figure}

An alternative, and slightly simpler, solution for the velocity reads as
\begin{eqnarray}
v^1&=&\frac{1}{2e_{33}} \left[\left({}^{(2,1,3)}C - 1 \right) A + e_{33}^2\right], \nonumber \\
v^2&=&-\frac{1}{2e_{33}} \left[-\left({}^{(1,2,3)}C - 1 \right) A + e_{33}^2\right], \nonumber 
\end{eqnarray}
\begin{eqnarray}
v^3&=&\frac{1}{2e_{22}} \left[\left({}^{(1,2,2)}C - 1 \right) A +e_{22}^2\right], \label{alternative_velocity_solution}
\end{eqnarray}
In other words, 4 clocks are necessary to determine all components of the linear velocity, see figure \ref{fig_4} for a symbolical sketch of the solution.

\subsection{Curvature determination}\label{paragraph_fermi_normal_curvature_determination}

Now we turn to the determination of the curvature in a general spacetime by means of clocks. We consider the non-vacuum case first, when one needs to measure 20 independent components of the Riemann curvature tensor $R_{abc}{}^d$. 

\begin{figure}
\begin{center}
\includegraphics[width=5.5cm,angle=-90]{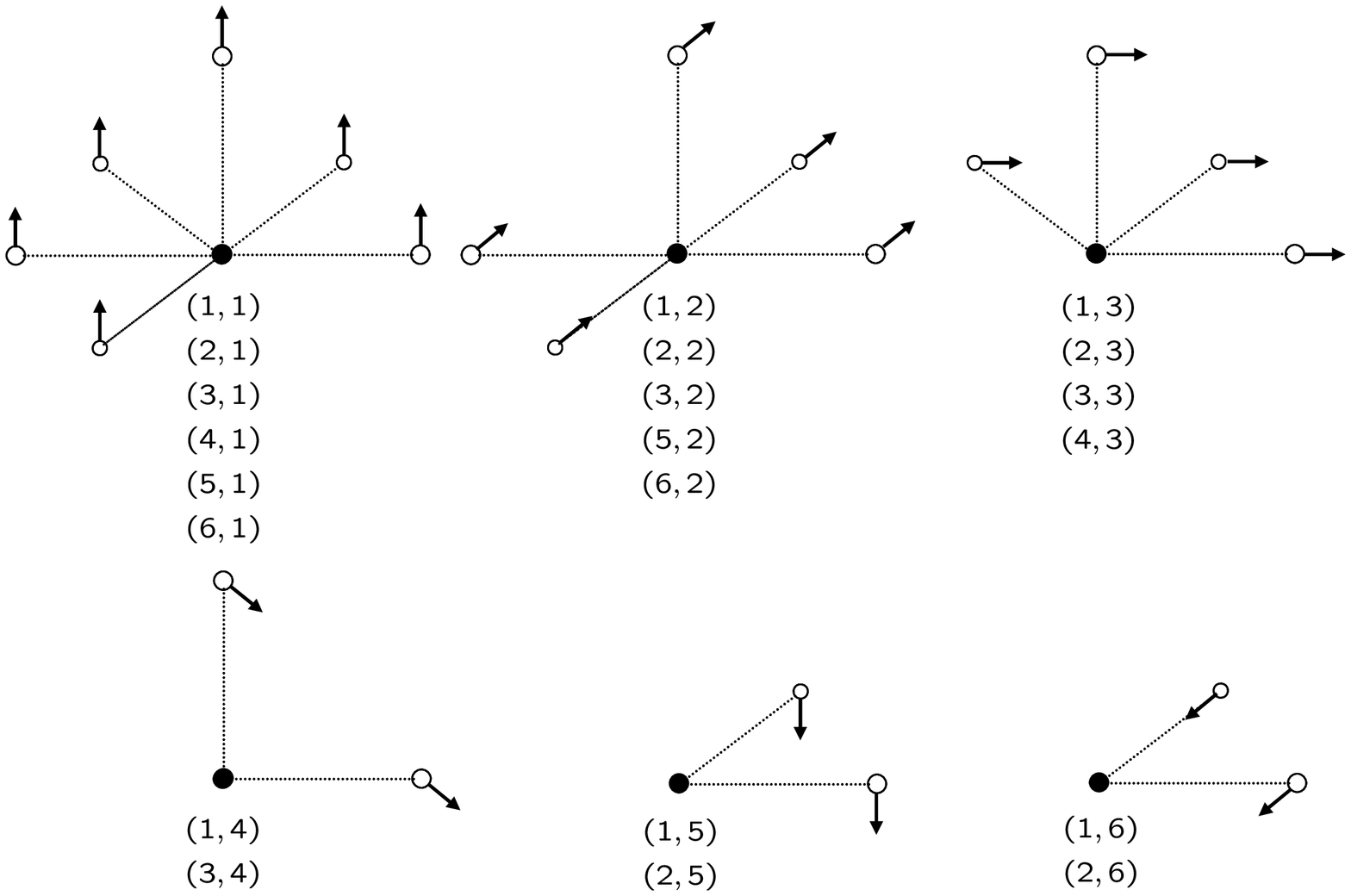}
\end{center}
\caption{\label{fig_5} Symbolical sketch of the explicit solution for the curvature (\ref{curvature_1})-(\ref{curvature_21}). In total 21 suitably prepared clocks (hollow circles) are needed to determine all curvature components. The observer is denoted by the black circle. Note that all ${}^{(1 \dots 6)}v^{a}$, but only ${}^{(1 \dots 3)}y^{a}$ are needed in the solution.}
\end{figure}

Again we start by rearranging the system (\ref{general_freq_ratio_definition}):
\begin{widetext}
\begin{eqnarray}
 {}^{(n)}y^\alpha  {}^{(n)}y^\beta \left(-R_{0 \alpha \beta 0} - \frac{4}{3} R_{\gamma \alpha \beta 0} \, {}^{(m)}v^\gamma + \frac{1}{3} R_{ \alpha \gamma \delta \beta } \, {}^{(m)}v^\gamma {}^{(m)}v^\delta \right) = B({}^{(n)}y^\alpha, {}^{(m)}v^\alpha, {}^{(p)}a^\alpha, {}^{(q)}\omega^\alpha), \label{curvature_master}
\end{eqnarray}
where
\begin{eqnarray}
&&B(y^\alpha, v^\alpha, a^\alpha, \omega^\alpha) := (1 - v^2) \left( C-1 \right) - 2 a_\alpha y^\alpha - y^\alpha y^\beta \left( a_\alpha a_\beta - \delta_{\alpha \beta} \omega_\gamma \omega^\gamma + \omega_\alpha \omega_\beta \right) - 2 v^\alpha  \varepsilon_{\alpha \beta \gamma}  y^\beta \omega^\gamma . \label{rhs_curvature_determination} 
\end{eqnarray}
Analogously to our analysis of the gravitational compass \cite{Puetzfeld:Obukhov:2016:1}, we may now consider different setups of clocks to measure as many curvature components as possible. The system in (\ref{curvature_master}) yields (please note that only the position and the velocity indices are indicated):
\begin{eqnarray}
01 : R_{1010}&=& {}^{(1,1)}B, \label{curvature_1} \\
02 : R_{2110}&=& \frac{3}{4}{c^{-1}_{22}c^{-1}_{42}\left(c_{22}-c_{42}\right)^{-1}} \left({}^{(1,1)}B c_{22}^2 - {}^{(1,1)}B c_{42}^2 + {}^{(1,2)}B c_{42}^2 - {}^{(1,4)}B c_{22}^2  \right) , \label{curvature_2} \\
03 : R_{1212}&=& -3 {c^{-1}_{22}c^{-1}_{42}\left(c_{22}-c_{42}\right)^{-1}} \left({}^{(1,1)}B c_{22} - {}^{(1,1)}B c_{42} + {}^{(1,2)}B c_{42} - {}^{(1,4)}B c_{22}  \right), \label{curvature_3} \\
04 : R_{3110}&=& \frac{3}{4}{c^{-1}_{33}c^{-1}_{63}\left(c_{33}-c_{63}\right)^{-1}} {\left({}^{(1,1)}B c_{33}^2 - {}^{(1,1)}B c_{63}^2 + {}^{(1,3)}B c_{63}^2 - {}^{(1,6)}B c_{33}^2  \right)}, \label{curvature_4} \\
05 : R_{1313}&=& -3 {c^{-1}_{33}c^{-1}_{63}\left(c_{33}-c_{63}\right)^{-1}}{\left({}^{(1,1)}B c_{33} - {}^{(1,1)}B c_{63} + {}^{(1,3)}B c_{63} - {}^{(1,6)}B c_{33}  \right)}, \label{curvature_5} \\
06 : R_{1213}&=& \frac{3}{2} c^{-1}_{52} c^{-1}_{53} \left(-\,{}^{(1,5)}B + R_{1010} - \frac{4}{3}R_{2110}c_{52} - \frac{4}{3}R_{3110} c_{53} - \frac{1}{3}R_{1212} c^2_{52}- \frac{1}{3}R_{1313} c^2_{53} \right), \label{curvature_6}\\
07 : R_{2020}&=& {}^{(2,2)}B, \label{curvature_7}\\
08 : R_{0212}&=& \frac{3}{4} c^{-1}_{11} \left({}^{(2,1)}B - R_{2020} +\frac{1}{3} R_{1212}c_{11}^2 \right), \label{curvature_8}\\
09 : R_{3220}&=& \frac{3}{4}{c^{-1}_{33}c^{-1}_{53}\left(c_{33}-c_{53}\right)^{-1}} {\left({}^{(2,2)}B c_{33}^2 - {}^{(2,2)}B c_{53}^2 + {}^{(2,3)}B c_{53}^2 - {}^{(2,5)} B c_{33}^2  \right)}, \label{curvature_9} \\
10 : R_{2323}&=& -3{c^{-1}_{33}c^{-1}_{53}\left(c_{33}-c_{53}\right)^{-1}}{\left({}^{(2,2)}B c_{33} - {}^{(2,2)}B c_{53} + {}^{(2,3)}B c_{53} - {}^{(5,2)}B c_{33}  \right)}, \label{curvature_10}
\end{eqnarray}
\begin{eqnarray} 
11 : R_{3212}&=& \frac{3}{2} c^{-1}_{61} c^{-1}_{63} \left( -\,{}^{(2,6)}B + R_{2020} + \frac{4}{3}R_{0212}c_{61}- \frac{4}{3} R_{3220} c_{63} - \frac{1}{3} R_{1212} c^2_{61}-\frac{1}{3}R_{2323}c^2_{63}\right), \label{curvature_11} \\
12 : R_{3030}&=& {}^{(3,3)}B, \label{curvature_12}\\
13 : R_{0313}&=& \frac{3}{4} c^{-1}_{11} \left({}^{(3,1)}B - R_{3030} +\frac{1}{3} R_{1313}c_{11}^2 \right), \label{curvature_13}\\
14 : R_{0323}&=& \frac{3}{4}c^{-1}_{22} \left({}^{(3,2)}B - R_{3030} +\frac{1}{3} R_{2323}c_{22}^2 \right), \label{curvature_14}\\
15 : R_{3132}&=& \frac{3}{2} c^{-1}_{41} c^{-1}_{42} \left(-\,{}^{(3,4)}B + R_{3030} + \frac{4}{3}R_{0313} c_{41} + \frac{4}{3} R_{0323} c_{42} - \frac{1}{3} R_{1313} c^2_{41} - \frac{1}{3}R_{2323}c^{2}_{42}\right), \label{curvature_15} \\
16 : R_{2010}&=& \frac{1}{2} \left({}^{(4,1)}B - R_{1010} - R_{2020} - \frac{4}{3} R_{0212}c_{11} - \frac{4}{3} R_{2110}c_{11} +\frac{1}{3} R_{1212}c^{2}_{11} \right), \label{curvature_16} \\
17 : R_{3020}&=& \frac{1}{2} \left({}^{(5,2)}B - R_{2020} - R_{3030} - \frac{4}{3}R_{0323}c_{22} - \frac{4}{3}R_{3220}c_{22}+\frac{1}{3}R_{2323}c^{2}_{22}\right), \label{curvature_17} \\
18 : R_{3010}&=& \frac{1}{2} \left({}^{(6,1)}B - R_{1010} - R_{3030} - \frac{4}{3} R_{0313}c_{11} - \frac{4}{3} R_{3110}c_{11}+\frac{1}{3} R_{1313}c^{2}_{11}\right). \label{curvature_18}
\end{eqnarray}
Introducing the abbreviations 
\begin{eqnarray}
K_1&:=& \frac{3}{4} c_{33}^{-1}\left[-\,{}^{(4,3)}B+R_{1010} + 2R_{2010} +R_{2020} - \frac{4}{3}(R_{3110} + R_{3220})c_{33}-\frac{1}{3}(R_{1313} + 2R_{3132} + R_{2323})c^2_{33} \right],\label{k1_definition} \\
K_2&:=& \frac{3}{4}c_{11}^{-1}\left[-\,{}^{(5,1)}B + R_{2020} + 2R_{3020} + R_{3030} +\frac{4}{3}(R_{0212} + R_{0313})c_{11}-\frac{1}{3}(R_{1212} + 2R_{1213} + R_{1313})c^2_{11} \right],\label{k2_definition} \\
K_3&:=&\frac{3}{4} c_{22}^{-1}\left[-\,{}^{(6,2)}B + R_{1010} + 2R_{3010} + R_{3030} -\frac{4}{3}(R_{2110} + R_{0323})c_{22} - \frac{1}{3}(R_{1212} + 2R_{3212}+ R_{2323})c^2_{22} \right],\label{k3_definition}
\end{eqnarray}
\end{widetext}
we find the remaining three curvature components 
\begin{eqnarray}
19 : R_{1023}&=& \frac{1}{3} \left(K_3 - K_1\right) , \label{curvature_19}\\
20 : R_{2013}&=& \frac{1}{3} \left(K_2 - K_1 \right) , \label{curvature_20}\\
21 : R_{3021}&=& \frac{1}{3} \left(K_3 - K_2\right) . \label{curvature_21}
\end{eqnarray}
See figure \ref{fig_5} for a symbolical sketch of the solution. The $B$'s in these equations can be explicitly resolved in terms of the $C$'s
\begin{eqnarray}
{}^{(1,1)}B&=&\left( 1 - c_{11}^2 \right) \left( {}^{(1,1)}C-1 \right), \label{B11_curved_case} \\
{}^{(1,2)}B&=&\left( 1 - c_{22}^2 \right) \left( {}^{(1,2)}C-1 \right), \label{B12_curved_case} \\
{}^{(1,3)}B&=&\left( 1 - c_{33}^2 \right) \left( {}^{(1,3)}C-1 \right), \label{B13_curved_case} \\
{}^{(1,4)}B&=&\left( 1 - c_{41}^2 - c_{42}^2 \right) \left( {}^{(1,4)}C-1 \right), \label{B14_curved_case}\\
{}^{(1,5)}B&=&\left( 1 - c_{52}^2 - c_{53}^2 \right) \left( {}^{(1,5)}C-1 \right), \label{B15_curved_case} \\
{}^{(1,6)}B&=&\left( 1 - c_{61}^2 - c_{63}^2 \right) \left( {}^{(1,6)}C-1 \right), \label{B16_curved_case} 
\end{eqnarray}
\begin{eqnarray}
{}^{(2,1)}B&=&\left( 1 - c_{11}^2 \right) \left( {}^{(2,1)}C-1 \right), \label{B21_curved_case} \\
{}^{(2,2)}B&=&\left( 1 - c_{22}^2 \right) \left( {}^{(2,2)}C-1 \right), \label{B22_curved_case} \\
{}^{(2,3)}B&=&\left( 1 - c_{33}^2 \right) \left( {}^{(2,3)}C-1 \right), \label{B23_curved_case} \\
{}^{(2,5)}B&=&\left( 1 - c_{52}^2 - c_{53}^2 \right) \left( {}^{(2,5)}C-1 \right), \label{B25_curved_case} \\
{}^{(2,6)}B&=&\left( 1 - c_{61}^2 - c_{63}^2 \right) \left( {}^{(2,6)}C-1 \right), \label{B26_curved_case} \\
{}^{(3,1)}B&=&\left( 1 - c_{11}^2 \right) \left( {}^{(3,1)}C-1 \right), \label{B31_curved_case}\\
{}^{(3,2)}B&=&\left( 1 - c_{22}^2 \right) \left( {}^{(3,2)}C-1 \right), \label{B32_curved_case} \\
{}^{(3,3)}B&=&\left( 1 - c_{33}^2 \right) \left( {}^{(3,3)}C-1 \right), \label{B33_curved_case} \\
{}^{(3,4)}B&=&\left( 1 - c_{41}^2 - c_{42}^2 \right) \left( {}^{(3,4)}C-1 \right), \label{B34_curved_case}\\ 
{}^{(4,1)}B&=&\left( 1 - c_{11}^2 \right) \left( {}^{(4,1)}C-1 \right), \label{B41_curved_case}\\
{}^{(4,3)}B&=&\left( 1 - c_{33}^2 \right) \left( {}^{(4,1)}C-1 \right), \label{B43_curved_case} \\
{}^{(5,1)}B&=&\left( 1 - c_{11}^2 \right) \left( {}^{(5,1)}C-1 \right), \label{B51_curved_case} \\
{}^{(5,2)}B&=&\left( 1 - c_{22}^2 \right) \left( {}^{(5,2)}C-1 \right), \label{B52_curved_case} \\
{}^{(6,1)}B&=&\left( 1 - c_{11}^2 \right) \left( {}^{(6,1)}C-1 \right), \label{B61_curved_case} \\
{}^{(6,2)}B&=&\left( 1 - c_{22}^2 \right) \left( {}^{(6,2)}C-1 \right). \label{B62_curved_case} 
\end{eqnarray}

\subsection{Vacuum spacetime}\label{subsection_vacuum_spacetime}

\begin{figure}
\begin{center}
\includegraphics[width=5.5cm,angle=-90]{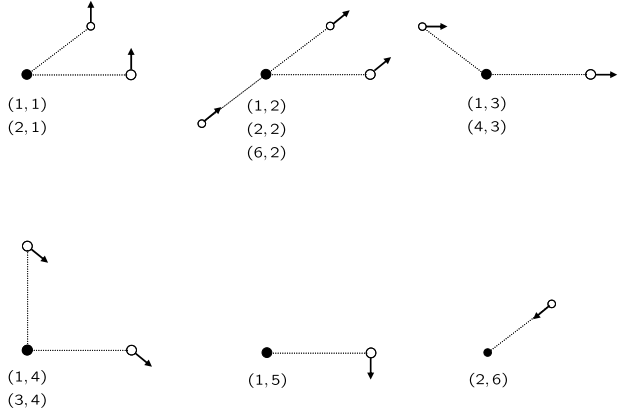}
\end{center}
\caption{\label{fig_6} Symbolical sketch of the explicit vacuum solution for the curvature (\ref{vacuum_curvature_1})-(\ref{vacuum_curvature_12}). In total 11 suitably prepared clocks (hollow circles) are needed to determine all curvature components. The observer is denoted by the black circle. Note that all ${}^{(1 \dots 6)}v^{a}$, but only ${}^{(1 \dots 3)}y^{a}$ are needed in the solution.}
\end{figure}

In vacuum the number of independent components of the curvature is reduced to the 10 components of the Weyl tensor $C_{abcd}$. Replacing $R_{abcd}$ in the compass solution (\ref{curvature_1})-(\ref{curvature_18}), and taking into account the symmetries of the Weyl tensor, we may use a reduced clock setup to completely determine the gravitational field. Note that all other components may be obtained from the double self-duality property $C_{abcd}=-\frac{1}{4}\varepsilon_{abef}\varepsilon_{cdgh}C^{efgh}$.
\begin{widetext}
\begin{eqnarray}
01 : C_{2323}&=& - {}^{(1,1)}B, \label{vacuum_curvature_1} \\
02 : C_{0323}&=& \frac{3}{4}{c^{-1}_{22}c^{-1}_{42}\left(c_{22}-c_{42}\right)^{-1}} \left({}^{(1,1)}B c_{22}^2 - {}^{(1,1)}B c_{42}^2 + {}^{(1,2)}B c_{42}^2 - {}^{(1,4)}B c_{22}^2  \right) , \label{vacuum_curvature_2} \\
03 : C_{3030}&=& 3 {c^{-1}_{22}c^{-1}_{42}\left(c_{22}-c_{42}\right)^{-1}} \left({}^{(1,1)}B c_{22} - {}^{(1,1)}B c_{42} + {}^{(1,2)}B c_{42} - {}^{(1,4)}B c_{22}  \right), \label{vacuum_curvature_3} \\
04 : C_{2020}&=& {}^{(2,2)}B, \label{vacuum_curvature_4} \\
05 : C_{3220}&=& \frac{3}{4}c^{-1}_{33} \left({}^{(1,3)}B + C_{2323} - \frac{1}{3} C_{2020}c_{33}^2 \right), \label{vacuum_curvature_5} \\
06 : C_{0313}&=& - \frac{3}{4}c^{-1}_{11} \left({}^{(2,1)}B -  C_{2020} - \frac{1}{3} C_{3030}c_{11}^2 \right), \label{vacuum_curvature_6} \\
07 : C_{3020}&=& - \frac{3}{2}c^{-1}_{52}c^{-1}_{53} \left({}^{(1,5)}B +  C_{2323} + \frac{4}{3} C_{0323}c_{52} - \frac{4}{3} C_{3220}c_{53} - \frac{1}{3} C_{3030}c^2_{52} - \frac{1}{3} C_{2020}c^2_{53}  \right), \label{vacuum_curvature_7} \\
08 : C_{3212}&=& - \frac{3}{2}c^{-1}_{61}c^{-1}_{63} \left({}^{(2,6)}B -  C_{2020} + \frac{4}{3} C_{0313}c_{61} + \frac{4}{3} C_{3220}c_{63} - \frac{1}{3} C_{3030}c^2_{61} + \frac{1}{3} C_{2323}c^2_{63}  \right), \label{vacuum_curvature_8} \\
09 : C_{3132}&=& - \frac{3}{2}c^{-1}_{41}c^{-1}_{42} \left({}^{(3,4)}B -  C_{3030} - \frac{4}{3} C_{0313}c_{41} - \frac{4}{3} C_{0323}c_{42} - \frac{1}{3} C_{2020}c^2_{41} + \frac{1}{3} C_{2323}c^2_{42}  \right). \label{vacuum_curvature_9}
\end{eqnarray}
With the abbreviations 
\begin{eqnarray}
K_1&:=& \frac{3}{4} c_{33}^{-1}\left[-{}^{(4,3)}B-C_{2323} + 2C_{3132} + C_{2020} + \frac{1}{3}(C_{2020} - 2C_{3132} - C_{2323})c^2_{33} \right],\label{k1_definition_2} \\
K_2&:=& \frac{3}{4}c_{11}^{-1}\left[-{}^{(5,1)}B + C_{2020} + 2C_{3020} + C_{3030} + \frac{1}{3}(C_{3030} - 2C_{3020} + C_{2020})c^2_{11} \right],\label{k2_definition_2} \\
K_3&:=&-\frac{3}{4} c_{22}^{-1}\left[{}^{(6,2)}B + C_{2323} - 2C_{3212} - C_{3030}  - \frac{1}{3}(C_{3030} - 2C_{3212}- C_{2323})c^2_{22} \right],\label{k3_definition_2}
\end{eqnarray}
\end{widetext}
the remaining three curvature components read
\begin{eqnarray}
10 : C_{1023}&=& \frac{1}{3} \left(K_3 - K_1\right), \label{vacuum_curvature_10}\\
11 : C_{2013}&=& \frac{1}{3} \left(K_2 - K_1\right), \label{vacuum_curvature_11}\\
12 : C_{3021}&=& \frac{1}{3} \left(K_3 - K_2\right). \label{vacuum_curvature_12}
\end{eqnarray}
A symbolical sketch of the solution is given in figure \ref{fig_6}.

\subsection{Constrained clock compass}\label{subsection_constrained_compass}

It is interesting to note that in the case of a constrained compass, when the relative velocities of the clocks to each other are vanishing, only six components of the curvature can be determined at best:
\begin{eqnarray}
01 : R_{1010}&=& {}^{(1)}B, \label{constrained_curvature_1} \\
02 : R_{2020}&=& {}^{(2)}B, \label{constrained_curvature_2} \\
03 : R_{3030}&=& {}^{(3)}B, \label{constrained_curvature_3} \\
04 : R_{2010}&=& \frac{1}{2} \left({}^{(4)}B - {}^{(1)}B - {}^{(2)}B\right), \label{constrained_curvature_4} \\
05 : R_{3020}&=& \frac{1}{2} \left({}^{(5)}B - {}^{(2)}B - {}^{(3)}B\right), \label{constrained_curvature_5} \\
06 : R_{3010}&=& \frac{1}{2} \left({}^{(6)}B - {}^{(1)}B - {}^{(3)}B\right). \label{constrained_curvature_6} 
\end{eqnarray}

\section{Conclusions and outlook}\label{sec_conclusions}

Here we proposed an experimental setup which we call a clock compass, in analogy to the usual gravitational compass \cite{Szekeres:1965,Puetzfeld:Obukhov:2016:1}. We have shown that a suitably prepared set of clocks can be used to determine all components of the gravitational field, i.e.\ the curvature, in General Relativity, as well as to describe the state of motion of a noninertial observer.  

We have worked out explicit clock compass setups in different situations, and have shown that in general 6 clocks are needed to determine the linear acceleration as well as the rotational velocity, while 4 clocks will suffice in the case of the velocity. Furthermore, we gave explicit setups which allow for a determination of all curvature components in general as well as in vacuum spacetimes by means of 21 and 11 clocks, respectively. In view of possible future experimental realizations it is interesting to note that restrictions regarding the choice of clock velocities in a setup lead to restrictions regarding the number of determinable curvature components. Further special cases should be studied depending on possible experimental setups.

In summary, we have shown how the gravitational field can be measured by means of an ensemble of clocks. Our results are of direct operational relevance for the setup of networks of clocks, especially in the context of relativistic geodesy. In geodetic terms, the given clock configurations may be thought of as a clock gradiometers. Taking into account the steadily increasing experimental accuracy of clocks, the results in the present paper should be combined with those from a gradiometric context, for example in the form of a hybrid gravitational compass -- which combines acceleration as well as clock measurements in one setup. Another possible application is the detection of gravitational waves by means of clock as well as standard interferometric techniques. An interesting question concerns the possible reduction of the number of required measurements by a combination of different techniques.

\begin{acknowledgments}
We thank Bahram Mashhoon for fruitful discussions and advice. This work was supported by the Deutsche Forschungsgemeinschaft (DFG) through the grant PU 461/1-1 (D.P.). The work of Y.N.O. was partially supported by PIER (``Partnership for Innovation, Education and Research'' between DESY and Universit\"at Hamburg) and by the Russian Foundation for Basic Research (Grant No. 16-02-00844-A). We also thank the DFG funded Research Training Group 1260 ``Models of Gravity'', as well as the Collaborative Research Center 1128 ``Relativistic Geodesy (geo-Q)''.
\end{acknowledgments}

\appendix

\section{Notations and conventions}\label{sec_notation}

\begin{table}
\caption{\label{tab_symbols}Directory of symbols.}
\begin{ruledtabular}
\begin{tabular}{ll}
Symbol & Explanation\\
\hline
$g_{a b}$ & Metric\\
$\sqrt{-g}$ & Determinant of the metric \\
$\delta^a_b$ & Kronecker symbol \\
$\varepsilon_{abcd}, \varepsilon_{\alpha\beta\gamma}$  & (4D, 3D) Levi-Civita symbol\\
$x^{a}$, $y^{a}$ & Coordinates \\
$s$, $\tau$ & Proper time \\
$\Gamma_{a b}{}^c$ & Connection \\
$R_{a b c}{}^d$, $C_{a b c}{}^d$ & Riemann, Weyl curvature \\
$\lambda_b{}^{(\alpha)}$ & (Fermi propagated) tetrad \\
$Y(s)$, $X(\tau)$ & (Reference) world line\\
$\xi^a$ & Constants in spatial Fermi coordinates\\
$v^\alpha$, $\omega^\alpha$, $V^\alpha$   & (Linear, rotational, combined) velocity\\
$a^\alpha$ & Acceleration\\
$b^\alpha$, $\eta^\alpha$ & Deriv. of (linear, rotational) acceleration \\ 
$C$ & Frequency ratio \\
$A$, $B$, $K_{1,2,3}$ & Auxiliary quantities\\
&\\
\hline
\multicolumn{2}{l}{{Operators}}\\
\hline
$\partial_i$, $\nabla_i$ & (Partial, covariant) derivative \\ 
$\frac{D}{ds} = $``$\dot{\phantom{a}}$'' & Total covariant derivative \\
$\frac{d}{ds} = $``$\stackrel{\circ}{\phantom{a}}$'' & Total  derivative \\
``$\bar{\phantom{A}}$'' & Power expansion  \\
&\\
\end{tabular}
\end{ruledtabular}
\end{table}

Our conventions for the Riemann curvature are as follows:
\begin{eqnarray}
&& 2 T^{c_1 \dots c_k}{}_{d_1 \dots d_l ; [ba] } \equiv 2 \nabla_{[a} \nabla_{b]} T^{c_1 \dots c_k}{}_{d_1 \dots d_l} \nonumber \\
& = & \sum^{k}_{i=1} R_{abe}{}^{c_i} T^{c_1 \dots e \dots c_k}{}_{d_1 \dots d_l} \nonumber \\
&& - \sum^{l}_{j=1} R_{abd_j}{}^{e} T^{c_1 \dots c_k}{}_{d_1 \dots e \dots d_l}. \label{curvature_def}
\end{eqnarray}
The Ricci tensor is introduced by $R_{ij} = R_{kij}{}^k$, and the curvature scalar is $R = g^{ij}R_{ij}$. The signature of the spacetime metric is assumed to be $(+1,-1,-1,-1)$. Latin indices run from $0, \dots, 3$, and Greek indices from $1, \dots, 3$.

\bibliographystyle{unsrtnat}
\bibliography{clocknet_bibliography}
\end{document}